\documentclass[aps,prl,reprint,showpacs,showkeys]{revtex4-1}
\usepackage{microtype}
\usepackage{amsthm}
\usepackage{amsmath}
\usepackage{amsfonts}
\usepackage[mathscr]{eucal}
\usepackage{amssymb,epsfig,setspace}
\usepackage{graphicx}
\usepackage{natbib}
\theoremstyle{plain}
\newtheorem{theorem}{Theorem}
\begin{document}
\title[Last word]{Comment on ``Note on the Analytical Solution of the
  Rabi Model''}%
\author{Andrzej J.~Maciejewski} \email{maciejka@astro.ia.uz.zgora.pl}
\affiliation{J.~Kepler Institute of Astronomy, University of Zielona
  G\'ora, Licealna 9, PL-65--417 Zielona G\'ora, Poland.}%
\author{Maria Przybylska}%
\email{M.Przybylska@proton.if.uz.zgora.pl} \affiliation{ Institute of
  Physics, University of Zielona G\'ora, Licealna 9, 65--417 Zielona
  G\'ora, Poland }%
\author{Tomasz Stachowiak} \email{stachowiak@cft.edu.pl}
\affiliation{%
  Center for Theoretical Physics PAS, Al. Lotnikow 32/46, 02-668
  Warsaw, Poland }%

\date{\today}%
\maketitle
The motivation of our recent preprint \cite{Maciejewski:12::x} was
paper ~\cite{Braak:11::,Braak:11:sup}.  It contains a very interesting
idea that enables to calculate the spectrum of the Rabi model.
However, this paper has two drawbacks. First of all, the presented
justification of the method contains an error.  Moreover, it has a
limited application to systems with a discreet symmetry.  This the
reason why we propose a new method in \cite{Maciejewski:12::x} which
can be applied to study more general class of systems than just the
Rabi model.  By no means was it our purpose to correct or improve
paper ~\cite{Braak:11::,Braak:11:sup}.

Let us explain the error contained in~\cite{Braak:11::,Braak:11:sup}.
The central role in this paper plays a certain function $G_+(x,z)$.
Here $x$ is the spectral parameter and $z$ is a complex variable. By
assumption this function is holomorphic is a certain neighbourhood of
the origin. As it is stated in~\cite{Braak:11:sup}, eq.~(16), $x$
belongs to the spectrum of the problem if and only if $G_+(x,z)$
vanishes for all $z\in\mathbb{C}$.  Just below equation (16)
in~\cite{Braak:11:sup}, the author wrote: \emph{Because $x$ is the
  only variable in $G_+(x; z) $ besides $z$, it suffices to solve
  $G_+(x; z) = 0$ for some \textbf{arbitrarily chosen} $z$}.  Clearly
it is not correct statement. This error was also noticed, and this is
not strange at all, by other authors, see, e.g., \cite{Moroz:12::a},
where the author writes: \emph{They (Braak's arguments) involved an
  ill motivated substitution and the argument that a sufficient
  condition for the vanishing of an analytic function $G_+(x; z)$
  (defined by eq. (16) of his supplement) for all $z \in\mathbb{C}$ is
  if it vanishes at a \textbf{single} point $z = 0$. The argument is
  essential to arrive at (41). However, as an example of any
  homogeneous polynomial shows, the argument is obviously invalid. }

To make this point clear beyond doubt, let us quote a standard theorem
of complex analysis \cite[p. 209]{Rudin:87::}
\begin{theorem}
  If $f$ and $g$ are holomorphic functions in a region $\Omega$ and if
  $f(z)=g(z)$ for all $z$ in some set which has a limit point in
  $\Omega$, then $f(z) = g(z)$ for all $z\in\Omega$.
\end{theorem}
In other words, for a holomorphic function to vanish in a region, one
needs to check it vanishes on an \textbf{infinite} set of points
having a limit point, not just on one point. Thus by choosing $z=0$
one merely checks the necessary, but not a sufficent condition as is
claimed in both~\cite{Braak:11:sup} and~\cite{Braak:12::b}. This
situation is only partially mitigated by the fact that $G_+(x,z)$
satisfies a second order linear homogeneous equation---see below.

In~\cite{Braak:12::b} the author says that \emph{our critique of his
  method is unfounded}.  As a matter of fact
in~\cite{Maciejewski:12::x} we did not criticise the general method
of~\cite{Braak:11::} as such. We simply point an evident error in the
very beginning of the justification/derivation of the method.  We did
not claim that it is impossible to correctly reformulate this method.
In~\cite{Braak:12::b} the author tries to correct his approach. Now
the necessary and sufficient condition for $x$ to belong to the
spectrum has the form of two equation (6a) and (6b) not just one as
equation (16) in ~\cite{Braak:11:sup}. Moreover, in
~\cite{Braak:12::b} it appears that the value of $z$ in those
conditions does matter. First of all, these additional conditions and
corrections of the original reasoning, are not to be found
in~\cite{Braak:11:sup}, and were presented as a reaction
to~\cite{Maciejewski:12::x}---this does not make the original article
retroactively correct, and its shortcomings remain.

Secondly, let us look in more detail at those additional equations
of~\cite{Braak:12::b}. As clarified, this could be rewritten as a
single (complex) equation when $\Im(z)\neq0$, because the series of
$G_+(x,z)$ has real coefficients, so $z^*$ supplies another point. The
key observation is that the function in question satisfies a second
order linear homogeneous equation so that we only need to make it
equal to zero at two \textbf{distinct} points.  (This step or
reasoning is crucial when one wants to avoid the direct application of
the quoted theorem, and is missing in both versions of the proof.)  In
case of equations (6a) and (6b) these are $z$ and $-z$. However, when
$z=0$ we only make the function zero at one point. Its not an
advantage, that the conditions are the same---rather we loose one of
them and need to supply one more in order to ensure that $G_+$
vanishes identically.  The reasoning that it is enough to check condition (6a) for just one $z$ requires $\Im (z)\neq 0$, which  obviously breaks down at $z=0$.  Numerical work seems to suggest that the condition at
zero is somehow distinguished, but in ~\cite{Braak:12::b} the
reasoning between (5) and (6) is one way, and only shows the necessary
condition.

In short, we still claim that there is no actual complete proof that
zeros of $G_{\pm}(x)$, i.e. when $z=0$, are necessary and
\textbf{sufficient} conditions for $x$ belongs to the spectrum.


In the end on his paper the author of~\cite{Braak:12::b} says:
\emph{The results of \cite{Maciejewski:12::x} neither correct nor
  extend the findings of ~\cite{Braak:11::} in these two cases, the
  only examples for which [2] presents explicit calculations.} In
fact, we neither corrected nor extended the method
of~\cite{Braak:11::} because it is still waiting for the main
correction.  Instead we provided a new, more general approach that
works for system without discreet symmetry, or with complex
coefficients simply by construction. It is also worth mentioning, that
our approach gives just one (real) equation where the other one offers
two (or one complex) and four equations in the generalised case.



\def\polhk#1{\setbox0=\hbox{#1}{\ooalign{\hidewidth
  \lower1.5ex\hbox{`}\hidewidth\crcr\unhbox0}}} \def\cprime{$'$}
  \def\cydot{\leavevmode\raise.4ex\hbox{.}} \def\cprime{$'$}

\end{document}